\documentclass[sigconf, nonacm, pdfa]{acmart}

\newif\ifrevisionmode
\revisionmodefalse

\newcommand\vldbdoi{10.14778/3828612.3828640}
\newcommand\vldbpages{2894 - 2907}
\newcommand\vldbvolume{19}
\newcommand\vldbissue{10}
\newcommand\vldbyear{2026}
\newcommand\vldbauthors{\authors}
\newcommand\vldbtitle{\shorttitle} 
\newcommand\vldbavailabilityurl{https://github.com/maxi-k/btrlog}
\newcommand\vldbpagestyle{empty} 

\usepackage[a-2b,mathxmp]{pdfx}[2018/12/22] 
\usepackage{xspace}
\usepackage{booktabs}
\usepackage{graphicx}
\usepackage[table,xcdraw]{xcolor}
\usepackage{colortbl}
\usepackage{enumitem}
\usepackage{float}
\definecolor{area0}{HTML}{2E3440}
\definecolor{sat0}{HTML}{202A40}
\definecolor{area1}{HTML}{D8DEE9}
\definecolor{sat1}{HTML}{CAD5E9}
\definecolor{area2}{HTML}{8FBCBB}
\definecolor{sat2}{HTML}{6BBCBA}
\definecolor{area3}{HTML}{88C0D0}
\definecolor{sat3}{HTML}{4EB3D0}
\definecolor{area4}{HTML}{81A1c1}
\definecolor{sat4}{HTML}{4E87C1}
\definecolor{area5}{HTML}{5E81AC}
\definecolor{sat5}{HTML}{205FAC}
\definecolor{area6}{HTML}{BF616A}
\definecolor{sat6}{HTML}{BF1626}
\definecolor{area7}{HTML}{D08770}
\definecolor{sat7}{HTML}{D04D23}
\definecolor{area8}{HTML}{EBCB8B}
\definecolor{sat8}{HTML}{EBB13E}
\definecolor{area9}{HTML}{A3BE8C}
\definecolor{sat9}{HTML}{8DBE64}
\definecolor{area10}{HTML}{B48EAD}
\definecolor{sat10}{HTML}{B470A7}
\definecolor{area11}{HTML}{469800}
\definecolor{sat11}{HTML}{469800}

\PassOptionsToPackage{capitalize,noabbrev,nameinlink}{cleveref}
\usepackage{cleveref}

\usepackage[nolist,nohyperlinks]{acronym}
\begin{acronym}
    \acro{db}[DB]{database}
    \acro{dbms}[DBMS]{database management system}
    \acro{lsn}[LSN]{log sequence number}
    \acro{rtt}[RTT]{round-trip}
    \acro{os}[OS]{operating system}
    \acro{oltp}[OLTP]{online transaction processing}
    \acro{olap}[OLAP]{online analytical processing}
    \acro{sw}[SW]{software}
    \acro{vm}[VM]{virtual machine}
    \acro{wal}[WAL]{write-ahead logging}
\end{acronym}

\newcommand{\myparagraph}[1]{\noindent\textbf{#1}.\xspace}
\newcommand{\btrlog}{\textsc{BtrLog}\xspace}

\usepackage{tikz}
\newcommand{\circled}[2][white]{\texttt{\textbf{\tikz[baseline=(char.base)]{
      \node[shape=circle,draw,inner sep=.5pt,fill=#1] (char) {%
        #2%
      }}}}}
\newcommand{\stepref}[1]{\circled[area8]{#1}}
\newcommand{\happystep}[1]{\circled[area8]{#1}}
\newcommand{\sadstep}[1]{\circled[area10]{#1}}

\usepackage[binary-units=true,per-mode=symbol]{siunitx}
\newcommand{\us}[1]{\qty{#1}{\micro\second}}

\newcommand{\ms}[1]{\qty{#1}{\milli\second}}
\newcommand{\kib}[1]{\qty{#1}{\kibi\byte}}
\newcommand{\gib}[1]{\qty{#1}{\gibi\byte}}
\newcommand{\mib}[1]{\qty{#1}{\mebi\byte}}
\newcommand{\gbit}[1]{\qty{#1}{\giga\bit\per\second}}
\usepackage{algorithm}
\usepackage{algpseudocode}

\newcommand{\pointref}[1]{\includegraphics{figures/point-#1.pdf}}

\ifrevisionmode
\newcommand{\marginnote}[1]{\if@firstcolumn\reversemarginpar\else\normalmarginpar\fi\marginpar{#1}}
\newenvironment{highlight}[1][]{\HLnote{#1}\color{blue}}{}
\newcommand{\revision}[2][]{\ifthenelse{\equal{#1}{}}{{\color[rgb]{0,0,0.8}#2}}{{\marginnote{\textit{\color[rgb]{0,0,0.8}#1}}\color[rgb]{0,0,0.8}#2}}}
\else
\newcommand{\marginnote}[1]{}

\newcommand{\revision}[2][]{#2}
\fi

\begin{document}
\title{\btrlog: Low-Latency Logging for Cloud Database Systems}

\author{Maximilian Kuschewski}
\affiliation{%
  \institution{Technische Universität München}
}
\email{maximilian.kuschewski@tum.de}

\author{Lam-Duy Nguyen}
\affiliation{%
  \institution{Technische Universität München}
}
\email{lamduy.nguyen@tum.de}

\author{Matthias Jasny}
\affiliation{%
  \institution{Technische Universität Darmstadt}
}
\email{matthias.jasny@tu-darmstadt.de}

\author{Tobias Ziegler}
\authornote{Work done while at Technische Universität München and Technische Universität Darmstadt.}
\affiliation{%
  \institution{TigerBeetle}
}
\email{tobias@tigerbeetle.com}

\author{Viktor Leis}
\affiliation{%
  \institution{Technische Universität München}
}
\email{leis@in.tum.de}

\author{Muhammad El-Hindi}
\affiliation{%
  \institution{Technische Universität München}
}
\email{muhammad.el-hindi@tum.de}

\begin{abstract}
Cloud database systems cannot rely on instance-local disks for write-ahead logging (WAL) durability, forcing WAL onto remote storage.
Existing options are unsatisfying: remote block storage like EBS is easy to adopt but adds substantial write latency and cost, while object storage offers excellent durability and low storage cost but is impractical for OLTP due to high latency and per-append cost.
Many cloud-native databases, therefore, depend on purpose-built logging backends, which are typically proprietary and tightly coupled to engine-specific replication and recovery protocols, limiting reuse.
We present \btrlog, a reusable cloud logging service that combines low-latency durable appends with low-cost archival for the common single-writer architecture.
\btrlog replicates log records across a quorum of SSD-backed log nodes in a single network round trip, reducing sensitivity to stragglers in commit latency.
To minimize storage cost, log nodes archive records to object storage as large segments, which are written asynchronously and off the latency-critical write path.
In our evaluation, \btrlog achieves lower latency than EBS and enables higher end-to-end transaction throughput when integrated into a DBMS\@.

\end{abstract}

\maketitle

\pagestyle{\vldbpagestyle}
\begingroup\small\noindent\raggedright\textbf{PVLDB Reference Format:}\\
\vldbauthors. \vldbtitle. PVLDB, \vldbvolume(\vldbissue): \vldbpages, \vldbyear.\\
\href{https://doi.org/\vldbdoi}{doi:\vldbdoi}
\endgroup
\begingroup
\renewcommand\thefootnote{}\footnote{\noindent
This work is licensed under the Creative Commons BY-NC-ND 4.0 International License. Visit \url{https://creativecommons.org/licenses/by-nc-nd/4.0/} to view a copy of this license. For any use beyond those covered by this license, obtain permission by emailing \href{mailto:info@vldb.org}{info@vldb.org}. Copyright is held by the owner/author(s). Publication rights licensed to the VLDB Endowment. \\
\raggedright Proceedings of the VLDB Endowment, Vol. \vldbvolume, No. \vldbissue\ %
ISSN 2150-8097. \\
\href{https://doi.org/\vldbdoi}{doi:\vldbdoi} \\
}\addtocounter{footnote}{-1}\endgroup

\ifdefempty{\vldbavailabilityurl}{}{
\vspace{.3cm}
\begingroup\small\noindent\raggedright\textbf{PVLDB Artifact Availability:}\\
The source code, data, and/or other artifacts have been made available at \url{\vldbavailabilityurl}.
\endgroup
}

\section{Introduction}\label{sec:introduction}\label{sec:intro}

\newcommand{\bad}{\cellcolor[HTML]{EA9999}\ensuremath{-}}
\newcommand{\good}{\cellcolor[HTML]{B6D7A8}$+$}
\newcommand{\meh}{\cellcolor[HTML]{FFE599}o}

\begin{table}
  \vspace{0.1cm}
  \caption{Comparison of storage options for WAL in the cloud.}%
  \label{tab:qualitative-cmp}
  \vspace{-0.2cm}
  \setlength{\tabcolsep}{2.8pt}
  \begin{tabular}{l@{\hspace{0.0pt}}rcccc}
    \toprule
                     &       & Availability & Latency   & \revision{Capacity/\$} & \revision{Append/\$} \\ \midrule
Local SSD            &       & \bad       & \good $+$ & \meh     & \good$+$                                      \\
EBS (io2)            & 1 AZ  & \good      & \meh      & \bad     & \meh                                          \\
S3 Express           & 1 AZ  & \good      & \bad      & \meh     & \bad                                          \\
BookKeeper           & 1 AZ  & \good      & \meh      & \bad     & \meh                                          \\
\textbf{\btrlog}     & 1 AZ  & \good      & \good     & \good    & \good                                         \\\midrule
S3                   & 3 AZs & \good $+$  & \bad $-$      & \good    & \bad                                          \\
\textbf{\btrlog}     & 3 AZs & \good $+$  & \meh      & \good    & \meh                                          \\\bottomrule 
\end{tabular}%
\small
\end{table}

\myparagraph{Durable cloud logging is remote}
On-premise database systems ensure durability through write-ahead logging (WAL) to a local disk or SSD\@.
In the cloud, instance-local storage is ephemeral and can be lost on instance failure or deprovisioning~\cite{inststoragepersistence}.
Cloud database systems must therefore store their log on durable remote storage.
An ideal cloud logging backend combines strong durability and availability with low-latency appends, low per-append cost, and low storage cost.
\Cref{tab:qualitative-cmp} qualitatively compares logging backends along these dimensions.

\myparagraph{Remote block storage: easy, but slow and expensive}
A straightforward approach is remote block storage, such as Amazon EBS, which provides a network-attached volume that can be re-attached to a different VM after failure.
EBS makes it straightforward to lift on-premise architectures to the cloud and is used in many deployments, including Amazon RDS~\cite{Barr_2009}.
However, even high-end EBS variants (e.g., io2) incur substantially higher write latency than instance-local SSDs and are expensive, with costs driven by both provisioned capacity and provisioned IOPS.

\myparagraph{Object storage: great archive, poor WAL backend}
Cloud object storage, such as S3, provides very high availability and durability, low cost per GB, and high throughput for scans.
These properties make it attractive as a log archive, and we use it for this purpose in our design.
However, placing a transactional WAL directly onto object storage (even its lower-latency variant, S3 Express) is impractical due to high latency and high per-append cost.

\myparagraph{Log systems exist, but are not reusable}
The limitations of general-purpose storage services for logging are well known, and many cloud-native database systems therefore rely on purpose-built logging backends.
Examples include Microsoft Socrates' XLOG~\cite{sigmod/AntonopoulosBDS19}, Huawei TaurusDB's PLog~\cite{sigmod/DepoutovitchCCL20}, and Neon's Safekeeper~\cite{NeonDBArchitectureDecisions}.
AWS's proprietary internal ``Journal'' service is used by Amazon Aurora DSQL, S3, DynamoDB, Lambda, and Kinesis~\cite{Brooker_2024}.
\revision[R1.W3]{
However, these systems are typically proprietary and tightly coupled to system-specific replication and recovery protocols, which makes them difficult to reuse outside their original architectures.
Because they are not publicly described in sufficient detail, the community still lacks both a reusable design and a systematic understanding of the latency/cost/availability tradeoffs for cloud logging backends.
}
In contrast, the open-source log system Apache BookKeeper~\cite{sigops/JunqueiraKR13} was designed for a pre-cloud world and does not exploit modern low-latency networks and object storage for WAL archival.

\myparagraph{\btrlog: fast quorum, cheap archive}
We argue that logging is a fundamental primitive and should be available as reusable infrastructure.
\btrlog~(pronounced ``better log'') is a cloud logging service that combines low-latency appends with low-cost archival.
It replicates log records across multiple SSD-backed log nodes; for the common single-writer architecture, each append sends one request per log node and becomes durable once a quorum (e.g., 2 of 3 nodes) has persisted it, limiting the impact of stragglers on commit latency.
Once a log segment (e.g., 16\,MB) fills, it is asynchronously written to S3, exploiting S3's low storage cost without suffering from its high latency.
\btrlog can be deployed within a single data center (Availability Zone or AZ in AWS) for the lowest latency and cost, or across AZs to protect the unarchived log tail against datacenter-wide failures.
As \Cref{tab:qualitative-cmp} shows, this design achieves high availability, low latency, and low per-append and storage cost.

\myparagraph{Novelty}
\revision[R1.W3]{
While specialized cloud logging backends already exist in the industry, they are proprietary and system-specific.
The novelty of \btrlog is therefore not merely to provide another logging service, but to make this design space explicit through a reusable service tailored to cloud WAL workloads.
}
We build on established distributed systems ideas, combining them in a way specifically tailored to cloud database logging.
To the best of our knowledge, \btrlog is the first cloud logging system that simultaneously
\begin{itemize}[leftmargin=2em]
\item achieves single network round-trip append latency for the common single-writer architecture,
\item provides a reusable service interface instead of coupling logging to a specific engine,
\item exploits cloud object storage for low-cost, high-durability archival without placing it on the critical path, and
\item is engineered for high-speed datacenter networks and NVMe SSDs, with explicit attention to tail latency and throughput.
\end{itemize}
Microbenchmarks in realistic AWS setups show that \btrlog reduces append latency by up to 4\texttimes{} compared to the high-end io2 variant of EBS, and end-to-end experiments with LeanStore~\cite{icde/LeisHK018} show how these gains translate into higher transaction throughput.

\section{Background: WAL in the Cloud}
\label{sec:background}

\begin{figure*}
  \centering
  \includegraphics[width=.99\linewidth]{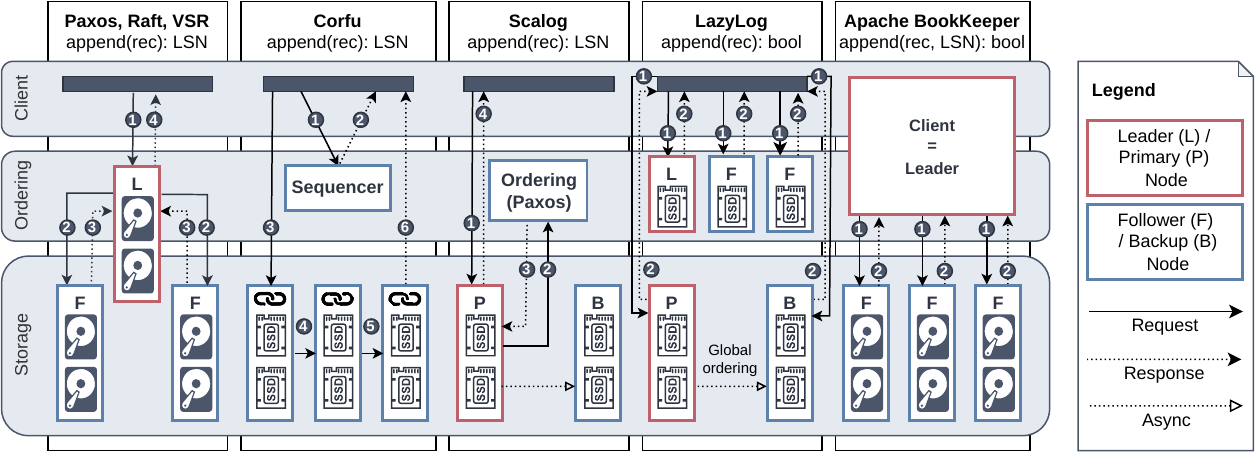}
  \caption{Comparison of existing log systems. Leader-based systems incur latency overheads due to centralized sequencing (Paxos, Corfu, Scalog). LazyLog and BookKeeper achieve optimal single-round-trip latency, which inspires \btrlog's design.}%
  \label{fig:shared_log_architectures}
\end{figure*}

In this section, we briefly summarize WAL semantics and the resulting requirements on a logging backend, then review existing log systems and cloud storage services.
The goal is to clarify the tradeoffs that motivate our design choices.

\subsection{WAL Semantics and Requirements}
\label{subsec:wal}

\myparagraph{Log interface}
Write-ahead logging (WAL) in database systems requires only a small interface.
During transaction execution, the DBMS appends log records via \texttt{append}~\cite{tods/MohanHLPS92}.
Each record is assigned a monotonically increasing \emph{log sequence number} (LSN).
At commit, the engine calls \texttt{sync} to ensure all records up to the commit LSN are persisted -- the ARIES paper~\cite{tods/MohanHLPS92} refers to this as \emph{forcing} the log.
For restart recovery, the engine issues a \texttt{scan} operation to retrieve log records starting from a given LSN.
During recovery or rollback, it calls \texttt{read} to fetch individual records by LSN for \emph{undo} processing.

\myparagraph{Key requirements}
Because the log is the source of truth for recovery, ARIES-style logging requires durable storage that survives process, system, and device failures~\cite{tods/MohanHLPS92}.
Accordingly, the logging backend must provide \textit{durability} and \textit{high availability}.
It must also support \textit{low-latency appends}, since WAL lies on the critical path of transaction processing.
Cloud database systems may also operate at extreme transaction rates (e.g., 70\,M/s~\cite{pvldb/Li19}), producing massive log volumes.
This makes \emph{cost-efficiency}, both per append and per GB, another important requirement.

\myparagraph{Single-writer semantics}
A key observation underlying our work is that many cloud and distributed databases achieve durability via per-partition (or per-replica) log streams.
Each log stream has an exclusive writer at any given time.
In primary/secondary architectures such as Amazon Aurora~\cite{sigmod/VerbitskiGSBGMK17}, Microsoft Azure SQL HyperScale~\cite{sigmod/AntonopoulosBDS19}, Alibaba PolarDB~\cite{pvldb/Li19}, and Google AlloyDB~\cite{alloydb}, only the writer instance issues updates and generates the corresponding log records.
Similarly, in strongly consistent distributed database systems such as Google Spanner~\cite{osdi/CorbettDEFFFGGHHHKKLLMMNQRRSSTWW12}, each replicated partition (Paxos group) elects a leader.
All writes are ordered through that leader, yielding a single total order of writes per group~\cite{osdi/CorbettDEFFFGGHHHKKLLMMNQRRSSTWW12}.
Even systems that expose multi-region write APIs (e.g., Cosmos DB~\cite{CosmosDB_2024}) implement durability via locally replicated physical partitions.
Ordering guarantees are defined by a leader within the associated replication scopes~\cite{CosmosDB_2024}.
As we discuss next, exploiting the monotonic, single-writer append order (via sequence numbers) is key to reducing WAL latency.

\subsection{Log Service Architectures}\label{subsec:logs_systems}

\myparagraph{Durability and ordering guarantees}
General-purpose dis\-tri\-bu\-ted log systems, and some distributed streaming systems (e.g., Apache Kafka), provide an append-only abstraction and use replication to achieve durability and availability.
In contrast to the single-writer WAL streams common in cloud database systems, they are designed for multiple concurrent writers.
They therefore include a \textit{sequencer} that orders writes across clients, linearizing appends.
The need to coordinate ordering and replication adds substantial latency on the append path, and architectures differ in how these responsibilities are
implemented, as \Cref{fig:shared_log_architectures} visualizes.

\myparagraph{Consensus-based systems (Paxos, Raft, VSR)}
As the left-hand side of \Cref{fig:shared_log_architectures} shows, leader-based consensus protocols~\cite{podc/OkiL88,usenix/OngaroO14,tocs/Lamport98} replicate a log across a set of nodes coordinated by a leader.
The leader serializes appends and replicates them to a quorum of followers, combining ordering and replication on the write path.
Each append requires four network hops (or two network round trips): client $\to$ leader and leader $\to$ follower.
The leader is both a throughput bottleneck and a source of latency variance.

\myparagraph{Throughput-optimized systems (Corfu, Scalog)}
Corfu~\cite{tocs/BalakrishnanMDPWW13}, Scalog~\cite{nsdi/0001CZLAR20}, and similar systems~\cite{osdi/BalakrishnanFSD20,sosp/JiaW21} decouple sequencing from storage.
This avoids the leader bottleneck by scaling the sequencer layer across nodes.
However, these systems still use \emph{eager ordering}~\cite{sosp/LuoBHAG24}: clients receive an LSN for an append once both ordering and replication are complete (\texttt{append(rec) $\rightarrow$ LSN}).
As~\Cref{fig:shared_log_architectures} shows, this results in additional network communication across layers.
The replication strategy further impacts latency: Corfu's chain replication requires six network hops; Scalog's primary-backup strategy reduces it to four, which is still well above the single-round-trip latency achievable with single-writer WAL semantics.

\myparagraph{Lazy ordering (LazyLog)}
LazyLog~\cite{sosp/LuoBHAG24} reduces append latency by separating ordering from durability.
Clients send metadata to sequencers while also sending data to storage nodes in parallel.
Appends are acknowledged without an LSN (\texttt{append(rec) $\rightarrow$ bool}) since ordering is resolved lazily during reads.
This yields single round-trip appends, but leaving LSNs unspecified at append time is incompatible with ARIES-style WAL, which requires a well-defined log order (and commit LSNs) on the write path.

\myparagraph{Single-writer systems (Apache BookKeeper)}
In contrast to multi-writer log systems, Apache BookKeeper~\cite{sigops/JunqueiraKR13} targets single-writer logs, as in primary/replica database architectures.
On writer failure (e.g., a crash), safe failover requires additional mechanisms.
BookKeeper therefore implements \emph{fencing} and \emph{recovery} to prevent split-brain scenarios during failover.
With a single writer, the order of records is implicitly defined by the writer's append sequence.
Thus, sequencing shifts to the client, eliminating the need for dedicated ordering nodes and their associated latency overheads.
BookKeeper uses client-driven quorum replication; in the common case, an append completes in a single client round trip.

\revision[]{
\myparagraph{Implications for \btrlog}
All discussed systems and architectures can provide the high durability and availability required for WAL in the cloud.
However, the previous discussion shows that architectural decisions directly affect the append latency of a logging system.
Similarly, cost efficiency is also impacted by architectural decisions such as the integration of cloud object storage.
The rest of the paper describes how \btrlog achieves low latency and cost while meeting WAL durability and availability requirements.
}

\section{\btrlog Design and Deployment}%
\label{sec:btrlog_design}

\myparagraph{Design space}
One possible solution for WAL in the cloud is writing log records directly to cheap and durable object storage.
However, object storage writes take tens of milliseconds~\cite{edbt/0001RJ0R25} and are expensive for workloads with many small writes -- such as WAL.
At the other extreme, one can envision a DRAM-based design similar to RAMCloud~\cite{tocs/OusterhoutGGKLM15}, where the client sends data to a quorum of nodes (``log nodes'') that keep data in DRAM and provide durability through replication.
This approach provides low latency but is expensive due to high DRAM cost~\cite{cidr/SteinertKL26}.
Furthermore, it is prone to correlated failures such as data center-wide outages or software bugs.
\revision[R3.W4]{
  Log nodes can mitigate correlated failures by writing data to local SSDs, which are guaranteed to persist data across power-loss reboots~\cite{inststoragepersistence}.
  In an extreme design, clients could persist data on remote SSDs directly using NVMe-over-TCP~\cite{nvme-over-tcp}, a protocol supported by the kernel and systems such as simplyblock~\cite{simplyblock} and DAOS~\cite{daos}.
  However, building a correct WAL backend using raw remote SSD semantics requires the client itself to handle log ownership, writer fencing for failover, remote node failures, and load balancing.
  \btrlog instead distributes this logic between a client library and protocol-aware log nodes, as described in the following.
}

\subsection{System Overview}%
\label{subsec:system_overview}

\begin{figure}
\centering
\includegraphics[width=\linewidth]{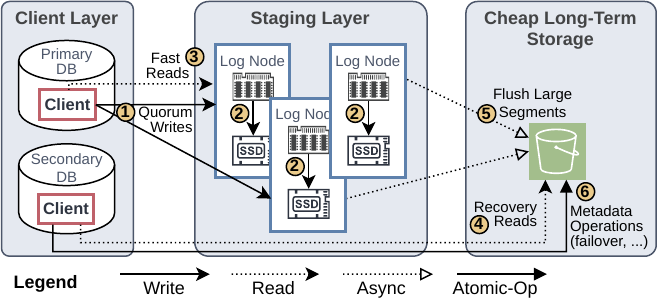}
\caption{System overview showing the append (write) and read paths across the staging layer and object storage.
}%
\label{fig:system_overview}
\end{figure}

\myparagraph{\btrlog's high-level design}
\btrlog overcomes durability and cost tradeoffs by combining a staging layer for low-latency appends with object storage for durable, cost-efficient persistence.
\btrlog stages appends on a cluster of log nodes and flushes them asynchronously to object storage in large segments.
Log nodes retain only a bounded unflushed \emph{tail} locally:
the tail resides in RAM for fast reads and is backed by local NVMe SSDs for recovery after crashes or power loss.
By flushing large segments asynchronously, \btrlog amortizes object store PUT cost and eliminates both object store latency and long-term storage capacity management.
\btrlog thus consists of four components, visualized in \cref{fig:system_overview}:
\begin{enumerate}[leftmargin=2em]
\item a \emph{client library} used by the DBMS to append log records,
\item a cluster of \emph{log nodes} that form the staging layer,
\item a \emph{cloud object store} for durable, low-cost storage, and
\item \emph{metadata storage} for log metadata and cluster configuration.
\end{enumerate}
\btrlog requires few metadata operations, which can be implemented using conditional writes (\emph{If-Match} in S3).
Using S3 or a compatible service for metadata simplifies deployment and avoids a separate coordination system such as ZooKeeper or etcd.

\myparagraph{Log abstraction and API}
\btrlog exposes a single-writer, append-only log abstraction.
The client library hides the internal storage tiering and provides a WAL-compatible API: \texttt{append}, \texttt{sync}, \texttt{scan}, and \texttt{read}.
Failover, metadata updates, and other control operations are handled transparently when the client \texttt{open}s a log for writing.

\myparagraph{Write path}
The append (write) path is \btrlog's hot path, optimized for single network round-trip commits.
As \Cref{fig:system_overview} illustrates, each client maintains a local LSN counter and replicates appends in parallel to all log nodes \stepref{1}.
Upon receiving an \texttt{append}, a node adds the record to an in-memory segment and writes it out-of-place~\cite{lee2026write} to a local NVMe SSD~\stepref{2}.
Only after the SSD write and fsync complete, the node acknowledges the append to the client.
The client library returns success to the DBMS after receiving acknowledgments from a quorum of nodes.
If multiple appends are in flight, the client library ensures that appends are acknowledged in LSN order to preserve ARIES semantics.
This client-driven sequencing eliminates a separate sequencer, enables single round-trip latency, and masks network and SSD latency jitter~\cite{pvldb/HaasLBL25} via quorum replication.

\myparagraph{Read path}
\btrlog distinguishes between hot and cold reads.
The hot, unflushed tail (e.g., for transaction rollback) is read from in-memory segments on log nodes~\stepref{3}.
Cold entries are read from flushed segments on object storage~\stepref{4} (e.g., for point-in-time recovery), leveraging its aggregate read bandwidth while reducing load on log nodes.
The client routes requests transparently via \texttt{read} and \texttt{scan} based on the requested LSN range and the replicated commit watermark, and may cache recently read tail records.

\subsection{Guarantees and Usage in Database Systems}
\label{subsec:guarantees}

\myparagraph{Durability and correlated failures}
\btrlog replicates each append to a quorum of $Q_w$ log nodes before acknowledging it and thus tolerates up to $Q_w - 1$ node failures without data loss.
For example, with 3 log nodes, $Q_w=2$, and the system tolerates 1 node failure.
Log segments are flushed asynchronously to object storage for long-term durability~\stepref{5}.
\revision[R1.W2]{%
  To tolerate correlated failures (e.g., software bugs or power outages), each node persists the hot tail on its local NVMe SSD, similar to prior work on highly durable storage backends~\cite{sosp/BornholtJACKMSS21}.
  This increases append latency, but strengthens durability:
  if all log nodes crash because of a software fault, in-memory data is lost, whereas data persisted on local SSDs can be recovered after fixing the fault and restarting the processes.
  }%

\myparagraph{Single writer, multiple readers}
\btrlog enforces a single-writer, multi-reader abstraction, supporting use cases like log shipping.
The writer maintains a commit watermark that is replicated across log nodes, allowing readers to distinguish committed from uncommitted data.
Enforcing a single writer prevents inconsistent appends during DBMS failover (e.g., when replacing a failed primary).

\myparagraph{Failure handling in the cloud}
Quorum replication masks transient node failures and network jitter, providing predictable tail latency for appends.
The client-side failover protocol~\stepref{6} prevents split-brain during DBMS failover.
\btrlog also handles compound and AZ+1 failure scenarios, as discussed in \Cref{sec:protocol}.

\myparagraph{Database integration}
As \Cref{fig:system_overview} illustrates, the DBMS integrates the \btrlog client into its WAL subsystem rather than writing log records to a storage device.
When the primary starts, or when explicitly requested, the client creates a new log on the log nodes and registers the log in the metadata management layer (c.f., \cref{sec:protocol}), which establishes the primary as the log's single writer.
It is then free to append log records as described in \cref{subsec:system_overview}.

\myparagraph{Transaction rollback}
The client may cache recently appended records in memory to enable low-latency reads for abort and rollback.
In addition, log nodes keep the hot log tail in memory, which accelerates hot reads.
The hot log tail is asynchronously flushed to object storage when it fills up or becomes cold.

\myparagraph{Database recovery}
The \btrlog client supports high-throughput reads of cold log data for database recovery.
For example, when a database node must replay a large log prefix to reconstruct database state, the client transparently fetches the hot tail from log nodes and streams older segments from high-bandwidth object storage~\cite{pvldb/DurnerL023}.

\myparagraph{Database failover}
For planned maintenance or primary failure, a backup node can take over as the writer.
When a \btrlog client opens an existing log for writing, it acquires ownership and fences off the previous writer to prevent conflicting appends.
\Cref{sec:protocol} details the protocol that implements these mechanisms.

\myparagraph{Ease of use}
Integrating \btrlog into an existing DBMS that relies on file system storage requires minor source code changes due to interface differences (\texttt{append} vs.~\texttt{write}).
Note, however, that file-system-based logging necessitates protecting against torn writes and identifying the last successful log record write during recovery.
\btrlog prevents both issues due to its record-level atomicity and failover support: its interface is purpose-built for database WAL.

\myparagraph{Parallel logging}
To avoid contention on a single log, some high-performance DBMSs support parallel logging through multiple (e.g., per-thread) log streams~\cite{sigmod/DiaconuFILMSVZ13,sigmod/HaubenschildS0L20,pvldb/XiaYPD20,pacmmod/NguyenAZL25}.
Such designs already manage transaction order across multiple log streams to ensure correct recovery and can thus simply use multiple \btrlog logs.

\subsection{Deployment Options}
\label{subsec:deployment_options}


\myparagraph{Configurable availability}
\btrlog supports two deployment options to adapt to different latency and durability/availability requirements: single-AZ and multi-AZ.
Although we use AWS terminology, the same concepts apply to other cloud providers as well.

\myparagraph{Single-AZ deployment}
To minimize append latency, \btrlog can be deployed in the same availability zone as the client application.
An AZ consists of one or more data centers within an AWS region, which are isolated from failures such as power, network outages, or flooding~\cite{availability-zones}.
Services such as EBS and S3 Express operate within a single AZ.
Intra-AZ network communication has latency of about \us{100} and is free of charge.
\btrlog log nodes reside in partitioned placement groups~\cite{placement_groups} to reduce correlated failures like rack-level power or network faults.
If the \textit{entire} AZ becomes unavailable (e.g., due to a major power outage), all log nodes in that AZ and \btrlog become unavailable as well.
Because \btrlog writes log records to instance-local SSDs, which persist data across power-loss reboots~\cite{inststoragepersistence}, data is still durable and can be recovered upon restart.
Records that were already flushed to object store remain accessible during the outage, since S3 replicates data across multiple AZs.

\myparagraph{Multi-AZ deployment}
To improve availability, \btrlog also supports deployments that span multiple AZs.
A straightforward approach would place three log nodes across three AZs, ensuring that AZ failures affect only one replica.
However, an additional independent failure (e.g., an SSD failure in a second AZ) can reduce a three-node deployment below the required 2/3 quorum, making the log unavailable for reads.
To tolerate such AZ$+1$ failures~\cite{sigmod/VerbitskiGSBGMK17}, \btrlog uses six nodes (two per AZ).
This configuration remains write-available after losing any two nodes or an entire AZ (4/6 append quorum), and remains read-available after losing up to three nodes (3/6 read quorum).
Even with six nodes, \btrlog persists records to local SSDs to guard against correlated software failures.

\myparagraph{Improving cost-efficiency in multi-AZ deployments}
Besides increasing latency, multi-AZ deployments also increase cost due to cross-AZ data transfer.
For example, replicating 100\,million 1\,KB log records from one AZ to another AZ of the same AWS region transfers \num{100}~GB and costs \$2.
The overhead is amplified by quorum replication:
if a client sends each append to all four remote nodes, it pays the transfer charge per remote replica (e.g., \(4 \cdot \$2 = \$8\)).
\btrlog reduces this overhead by 2\texttimes{} via hierarchical replication:
the client sends each record to one node per AZ, which forwards it to its peer in the same AZ.
This significantly reduces cost but only slightly increases latency because intra-AZ latency (about \us{100}) is significantly lower than cross-AZ latency (about \us{500}).

\section{Protocols and Fault Tolerance}
\label{sec:protocol}

Our design is simple: quorum-replicated appends to a staging layer, with asynchronous, segment-based persistence to object storage.
The challenge is to preserve \btrlog's guarantees under realistic cloud failure modes -- transient faults, correlated node/AZ outages, and DBMS failovers -- without sacrificing one-round-trip commits.
To this end, we employ a leader-based quorum protocol with view changes, similar to Apache BookKeeper, customized for \btrlog's object storage-oriented persistence path.
The protocol is grounded in two invariants: (i) log records commit in order, and (ii) no committed entries are lost, even in the presence of failures within our failure model.
We formalize these safety properties and an abstract model of the protocol in TLA+ and model-check them (see artifact URL).
The remainder of this section gives a high-level overview of the protocol design.

\myparagraph{Failure model}
\btrlog adopts the standard non-Byzantine crash-stop failure model in asynchronous networks~\cite{jacm/ChandraT96}.
Accordingly, our protocol tolerates message loss, reordering, duplication, delay, network partitions, log node failures, and client failures.
In the following, we discuss client and log node failures separately.
Client failures coincide with database failover.
Log node failures must not compromise the availability or durability of committed data in the log tail.
Because cloud object storage provides high durability and availability (e.g., S3 advertises 99.99\% availability and 11 nines of durability~\cite{aws_s3_durability}), we treat object storage and the metadata store as reliable and focus on failures in the staging and replication layers.

\myparagraph{Protocol overview}
\btrlog targets the same failure modes as Paxos-like protocols, but shifts responsibilities to reduce latency while preserving fault tolerance.
It assigns the leader/proposer role to the single writer (the client) to minimize commit latency, and leverages WAL monotonicity to avoid per-record ordering at log nodes.
\btrlog integrates object storage to simplify node recovery and avoid re-replication across log nodes.
Cloud object storage also serves as a highly available metadata store with atomic compare-and-swap (CAS) semantics for leader election (client failover).
In the following, we discuss how \btrlog uses these concepts to achieve fault tolerance.
We assume a single-AZ deployment for simplicity, but our explanations generalize to \btrlog's multi-AZ deployment.
\revision[]{
  We first describe the failure-free operations that enable low-latency appends and consistent reads, then we discuss failure scenarios.
}

\begin{figure}
\centering
\includegraphics[width=0.9\linewidth]{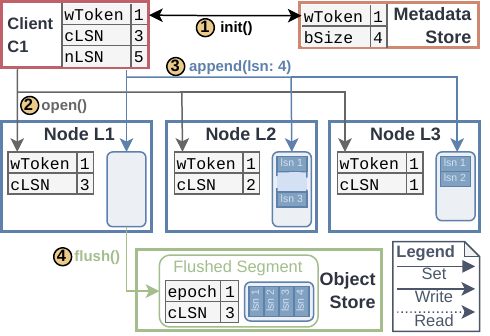}
  \vspace{-0.5em}
\caption{State for a single log without failures.
The client acquires a write token (\textit{wtoken}) from the metadata store and installs it on log nodes via \texttt{open}.
Appends are replicated, advancing the committed watermark (\textit{cLSN}).
Log segments are flushed to object storage in large (\textit{bSize}) segments.}%
\label{fig:happy_path}
\end{figure}

\subsection{Failure-Free Operation}\label{subsec:happy_path}

\myparagraph{Log creation}
Before appending, clients have to create a new log by initializing its metadata (such as the segment size (\textit{bSize})) in the metadata store, as shown in \cref{fig:happy_path} \happystep{1}.
To open a log for writing, the client acquires a write token (\textit{wtoken}) from the metadata store (c.f. \Cref{subsec:client_failure_handling}) and sends an \texttt{open} request to all log nodes to install that token \happystep{2}.
Before issuing append requests, the client waits for acknowledgments from a quorum of log nodes.

\myparagraph{Append operations}
The \btrlog client sends append requests to all log nodes and commits after replies from a write quorum \happystep{3}.
In addition to the payload, each append request includes three fields:
(1) the record's log sequence number (nLSN),
(2) the last committed LSN (cLSN, a commit watermark for readers), and
(3) the record's byte offset within the current segment, computed as the prefix sum of payload sizes of preceding records in the segment.
The prefix sum allows log nodes to place each record at a deterministic offset within a segment, enabling optimized flushing to object storage.

\myparagraph{Idempotent segment flushes}
On the failure-free fast path, log nodes flush full segments to object storage asynchronously \happystep{4}.
If all nodes wrote the same segment, this would incur 3\texttimes{} the PUT cost in a 3-node deployment.
To avoid redundant PUTs, \btrlog uses the prefix sum of record sizes to construct byte-by-byte identical segments across all nodes.
As a result, each node derives the same deterministic object name for the segment, making the flush idempotent.
Nodes create the object using a conditional PUT (with \emph{If-None-Match} in S3), so at most one concurrent uploader succeeds.
To optimize costs, a node checks whether the object already exists before issuing a PUT and skips the write if it does.
\revision[R3.D3]{%
  Since flushes are performed asynchronously, \btrlog can use expensive cryptographic hashing to avoid accidental hash collisions.
  In the rare case of packet loss or writer fencing, hashes may differ; \btrlog then tolerates duplicate flushes to reduce coordination overhead and performs data de-duplication during read operations.
}

\revision[\\R3.W1 R3.W2]{
\myparagraph{Read guarantees}
The read protocol guarantees that all committed data is readable. 
Internally, \btrlog distinguishes hot reads, served by log nodes, from cold reads, served by object storage.
The client handles both types of reads transparently to the DBMS.

\myparagraph{Hot data reads}
Hot data residing on log nodes is typically read either for transaction rollback (where records are read to undo a transaction), for log shipping, or for read-replication, as in Microsoft Socrates~\cite{sigmod/AntonopoulosBDS19}, for example.
Because the write path tolerates message loss, committed entries may be absent on some log nodes.
Similarly, the committed LSN watermark (cLSN) may be stale on nodes that have not yet received the latest appends.
Hot data reads are therefore performed as quorum reads, which enable clients to tolerate gaps on individual nodes and observe up-to-date committed LSN watermarks.
To optimize bandwidth utilization, readers may optimistically decide to read only from a single node and switch to quorum reads when encountering gaps or detecting a stale cLSN on that node.  
If the requested LSN range has already been flushed, log nodes redirect the request to object storage.

\myparagraph{Cold data reads}
\btrlog clients leverage object storage for high-throughput reads of cold data, as required for database recovery.
After a crash, the DBMS replays log segments directly from object storage, reducing load on log nodes and allowing them to prioritize low-latency tail operations.
To enable clients to build an index via LIST operations and use it to retrieve ranges, log nodes encode the log ID, epoch, and LSN range in addition to the hash of the data in the object key.
The index can be incrementally maintained on clients using metadata from log nodes (e.g., the last flushed LSN).

\myparagraph{Deferred eviction for low-latency reads}
The core protocol guarantees that no committed data is lost, but does not by itself guarantee low read latency for all access patterns.
For example, data requested for log shipping or transaction rollback may have already been evicted to object storage.
In practice, object store eviction can be controlled without modifying the core protocol using the \emph{LSN window} mechanism, which is described in \cref{sec:implementation}.
}

\begin{figure}
\centering
\includegraphics[width=0.9\linewidth]{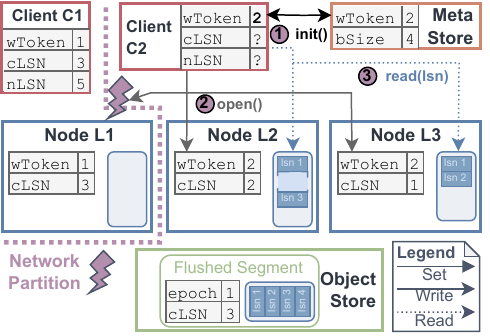}
  \vspace{-0.5em}
\caption{%
  Client failover: The new client acquires a write token (\textit{wtoken}) and installs it on a quorum of log nodes, fencing off the old writer.
  It repairs the tail by identifying the highest LSN and re-replicating records using the new \textit{wtoken}.%
}%
\label{fig:client_failure_handling}
\end{figure}

\subsection{Client Failover}
\label{subsec:client_failure_handling}

Given the client's leader role, client failures require explicit failover handling.
The key challenge is to preserve the single-writer invariant and prevent split-brain, in which two clients concurrently believe they hold leadership for the same log.
Without fencing, concurrent append operations can create divergent tails across log nodes and conflicting persisted segments in object storage.

\myparagraph{Running example}
We use the example depicted in \Cref{fig:client_failure_handling} to show how \btrlog prevents split-brain scenarios and enforces single-writer semantics through its failover protocol.
In this example, client C1 is isolated in a network partition and retains connectivity only to log node L1.
To replace the faulty client, a new client, C2, opens the existing log as the writer.

\myparagraph{Writer fencing}
Concurrent writes to the same log violate \btrlog's single-writer invariant.
We therefore use a monotonically increasing write token (\textit{wtoken}) to implement \emph{writer fencing}.
In \Cref{fig:client_failure_handling}, client C2 atomically increments the log's write token from 1 to 2 \sadstep{1} via an atomic metadata operation.
It then installs the token on log nodes by sending \texttt{open} and waiting for acknowledgments from a write quorum \sadstep{2}.
A log node accepts requests only for the highest token it has observed; if it has already seen a higher token, it rejects \texttt{open} and returns the higher value.
If the client cannot obtain a quorum of acknowledgments, it aborts the takeover and reports failure to the application (which may retry).

\myparagraph{Avoiding data loss}
Before taking over, a new writer must establish a correct log tail to preserve monotonicity and avoid losing committed data.
\Cref{fig:client_failure_handling} highlights a subtlety introduced by object storage:
Even if client C2 cannot reach log node L1, it may still observe segments that node L1 has already flushed to object storage.
Naively consulting object storage during failover would add substantial latency (tens of milliseconds per access), so the protocol avoids putting object storage reads on the critical path.

\myparagraph{Finding the log tail}
Fortunately, failover does not require using object storage.
To locate the tail of the log without losing committed data, the new writer determines the largest contiguous recoverable LSN prefix supported by a read quorum of log nodes.
To accelerate this step, log nodes piggyback their last committed LSN watermark when acknowledging the installation of the client's write token.
The client then starts reading log records sequentially \sadstep{3}, starting from the highest watermark it observed (cLSN = 2 in the example).

\myparagraph{Determining committed data}
The retrieved log record's LSN might be below the last committed watermark observed by a node, and only one copy of the record might exist.
A stricter setting that requires the writer to observe a quorum of copies might cause data loss, as illustrated in \Cref{fig:client_failure_handling}.
In the example, C2 observes only one copy of LSN 3 even though the previous writer replicated it to 2/3 nodes and reported success to the application (cLSN = 3 on C1).
Such cases occur because quorum replication can introduce gaps, e.g., due to message loss, and the last committed watermark can lag, especially when multiple appends are in flight.
To avoid data loss, new writers are conservative when inferring the log tail: they include every consecutive LSN present on at least one node in the quorum.
Only LSNs that a quorum of log nodes confirms as absent could not have been committed.
In the example, although LSN 4 is durably stored in object storage, C2 can infer it was not committed because a read quorum confirms it absent.

\myparagraph{Log repair and appending}
Under-replicated records (e.g., LSNs 2 and 3 in \cref{fig:client_failure_handling}) are more susceptible to data loss during network partitions.
For example, the only accessible copy of LSN 2 is lost when node L3 fails.
To improve durability of under-replicated records, the new client re-replicates such records to all nodes using the new write token \sadstep{4}.
These repair writes may fail if the new client is also fenced, in which case it stops the failover process and informs the application to handle retries.
Writer fencing can happen at any stage, but the protocol guarantees that concurrent repairs will recover the same or a higher log tail.
Once fencing and repair are complete, the new client continues appending to the same log.

\myparagraph{Reconstructing the client's log state}
As described earlier, the failover protocol determines all required metadata for the new writer's state, including the last committed LSN and the next LSN (nLSN).
During tail recovery, the writer also retrieves each record's byte offset from log nodes.
This information is used to reconstruct the segment prefix sum for idempotent object storage flushes.

\subsection{Node Failures}
\label{subsec:node_failures}

\myparagraph{Detecting node failures}
Although \btrlog uses a client-driven replication protocol, it does not depend on clients to detect or handle node failures.
If failure detection were delegated to clients, node outages could go unnoticed when a client idles or crashes, leaving the tail unreplicated and increasing the risk of data loss.
Instead, \btrlog detects node failures via peer-to-peer heartbeats between log nodes.
If a node misses heartbeats from a peer for a configurable interval, it marks that peer as failed.

\myparagraph{Handling node failures}
Traditional protocols, like VSR~\cite{podc/OkiL88}, Raft~\cite{usenix/OngaroO14}, Corfu~\cite{corfu_repo}, or Scalog~\cite{nsdi/0001CZLAR20}, use re-replication from surviving nodes to recover and rebuild failed nodes.
Healthy nodes serve read requests to recovering nodes, increasing load.
To avoid this overhead, \btrlog flushes a snapshot of the log segments to object storage rather than re-replicating the data to other nodes.
The snapshot flush persists all LSNs up to the point at which the failure was detected.
Entries committed \emph{after} detection are still replicated to a quorum by design.
Besides reducing load on healthy nodes, this strategy also enables new or recovering nodes to accept new writes immediately after receiving the current write token.

\myparagraph{Cost-efficient failure handling}
Flushing log segment snapshots to object storage on every node failure might, however, increase costs due to the number of PUT requests.
To reduce the flush costs, \btrlog introduces two optimizations:
First, log nodes trigger a flush only when they detect $N-Q_w$ node failures.
For example, in a cluster with 3 nodes and $Q_w=2$, a single failure would already trigger a flush.
In a cluster of 6 nodes and $Q_w=4$, snapshot flushes are only triggered after two failures.
Second, we deterministically assign a per-log flush leader: the leader flushes immediately, while other nodes first check whether the object already exists before writing.
This optimization avoids unnecessary flushes.

\myparagraph{Node replacement and reconfiguration}
\btrlog does not automatically redeploy failed nodes and assumes that a higher-level monitoring component handles such logic.
Currently, faulty nodes can be replaced by removing a node and replacing it with a new node, reusing the same DNS name or IP address.
More automated node replacement strategies and cluster resizing through reconfiguration are interesting avenues of future work.

\subsection{Unreliable Networks}
\label{subsec:unreliable_networks}

\myparagraph{Impact of network issues}
\btrlog assumes unreliable networks where append requests may get reordered or dropped entirely.  
Network issues can introduce gaps in node-local segments and delay the propagation of the committed watermark from the client to the log nodes.
Nodes may also flush incomplete or overlapping segments to object storage, potentially leading to uncommitted or duplicated records stored there.
For instance, a node may flush while client failover is in progress.
\Cref{fig:client_failure_handling} illustrates how this failure case can cause data to appear on object storage even without being replicated and committed by quorum.

\myparagraph{Handling uncommitted data}
To keep the write path fast, \btrlog postpones handling such cases to the read path.
A key invariant for the read path is that data in object storage is only \emph{durable}, not necessarily \emph{committed}.
Readers identify committed data through the last committed watermark (cLSN) that the client propagates to log nodes, which in turn include it as object metadata on each flush.
For example, when clients retrieve data with the \texttt{read} API, nodes return both the requested data and the committed LSN they observed.
This mechanism enables readers to distinguish committed data from merely durable data.
Similarly, when retrieving data from object storage, clients evaluate the committed watermark in the object's metadata to filter out uncommitted data if required.

\myparagraph{Handling overlapping data}
Due to quorum replication, log segments may have gaps and overlapping ranges.
For instance, some nodes might observe LSN1 and LSN3, while others receive LSN2 and LSN4.
When such overlapping log segments are flushed to object storage, e.g., due to a node failure, clients may observe overlapping copies of those segments.
The \btrlog client transparently merges log segments while reading from object storage.

\myparagraph{Handling duplicate data}
A more subtle issue can occur during client failover.
As \Cref{fig:client_failure_handling} illustrates, pending requests from an old writer (client C1) can fill up log segments and trigger a flush to object storage.
When a new writer (client C2) takes over the log, it might append new data with the same LSN, for instance, LSN4 in \Cref{fig:client_failure_handling}.
Once the new log segment is flushed to object storage, readers would observe multiple different copies of LSN4.
To resolve such conflicts in object storage, \btrlog utilizes the write token as an epoch.
Each flushed object encodes the epoch in its key and includes it in its metadata.
Similar to the last committed watermark, \btrlog clients use this information to ignore outdated data from prior epochs while streaming from object storage.

\section{Engineering \btrlog for Low Latency}%
\label{sec:impl}\label{sec:implementation}

\begin{figure}
  \centering
  \includegraphics[width=0.95\linewidth]{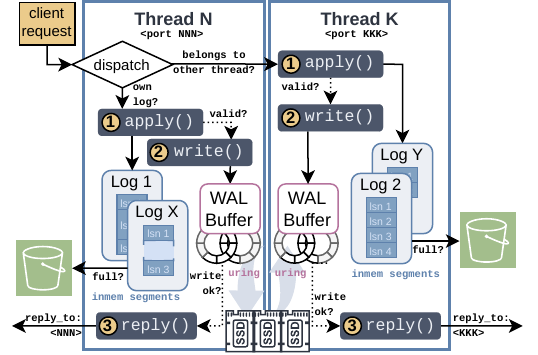}
  \caption{%
    \btrlog log node architecture for managing parallelism across requests, CPU cores, and SSDs.
  }%
  \label{fig:btrlog_impl}
\end{figure}

\myparagraph{Implementation overview}
Achieving high throughput and low latency requires efficient protocols and an optimized implementation that can effectively utilize modern networks, NVMe SSDs, and multi-core CPUs~\cite{DBLP:conf/damon/JasnyE0B25}.
Our prototype implements the latency-critical, failure-free append path; failure cases are modeled in TLA+.

\myparagraph{Lower latency bound}
Microbenchmarks using \textit{sockperf}~\cite{sockperf} and \textit{fio}~\cite{fio_repo} in AWS show that a cluster of \textit{c6id.metal} instances can achieve approximately \us{40} network round-trip and \us{30} SSD write latency without load in a partitioned placement group.
Our implementation thus targets \us{70} latency without load.
Achieving this requires careful optimization of cache misses, allocations, and system calls:
For example, system calls and large allocations can take hundreds of nanoseconds or even microseconds, potentially deteriorating response latency by several percent.
Thus, our technology stack needs to allow for fine-grained control over such operations.

\myparagraph{Technology stack: Rust, io\_uring, UDP}
\btrlog is implemented in Rust.
Since no tested off-the-shelf asynchronous Rust runtime offered consistently low latency and observability, we implemented our own.
Our custom asynchronous runtime is built on \textit{io\_uring}~\cite{linux-uring}, which allows batching system calls for common operations such as \textit{send}, \textit{recv}, and \textit{write}~\cite{DBLP:journals/corr/abs-2512-04859}.
Finally, since the quorum-based \btrlog protocol obviates the need for many of TCP's guarantees, we use UDP by default, avoiding stream and connection overhead.

\myparagraph{Multi-Core architecture}
\btrlog is designed as a multi-tenant cloud service, so it expects to manage many independent open log streams simultaneously.
Conversely, each log requires sequential semantics internally, and the most common \textit{append} operation needs few CPU cycles:
As \cref{fig:btrlog_impl} illustrates, each thread \stepref{1} checks the message and copies it to the in-memory log segment if valid, \stepref{2} asynchronously writes it to the local SSD, ensuring fsync semantics, \stepref{3} replies to the client once data is durable, and then asynchronously flushes full segments to S3 if necessary.
This model naturally yields an architecture where each log is owned by a single thread that sequentially processes requests for that log.

\myparagraph{Per-Thread runtime}
I/O dominates the end-to-end latency of requests, so I/O wait times should be utilized to process other requests.
One could achieve this by using synchronous I/O requests and letting the OS schedule other threads.
However, the resulting thread-to-core oversubscription and excessive OS scheduling can lead to latency spikes.
\btrlog thus spawns one thread per CPU core and uses cooperative scheduling within a custom asynchronous Rust runtime to interleave I/O and CPU work on each thread.

\myparagraph{Symmetric networking}
Many systems use dedicated networking threads to simplify the architecture and guarantee fast packet acceptance.
However, this asymmetric approach requires cross-thread synchronization for every packet and is detrimental on small nodes with few CPUs, since one CPU thread is designated to networking only.
In BtrLog, every thread accepts network requests on its own port using the \textit{io\_uring}-based runtime and processes them directly.

\myparagraph{Internal load balancing}
The symmetric networking approach has a significant drawback:
If clients send requests directly to individual threads on \btrlog nodes, the node cannot balance log segments across threads.
Load balancing is important since we expect multi-tenant workloads to serve tenants with both very high and very low append frequency.
Having multiple high-frequency logs handled by the same thread would negatively impact peak throughput.
To enable load balancing while still allowing threads to respond to clients directly and reduce cross-thread synchronization, our network protocol includes a \texttt{reply\_to} field that specifies a port number.
Clients can use \textit{any} previously used port, including a well-known port for initial requests, and the \btrlog node may internally forward the message to a different thread, handing over the connection.
The receiving thread responds to the client directly, setting its own port in the \texttt{reply\_to} field, which the client uses for future requests -- until the log moves to a different thread again.
\cref{fig:btrlog_impl} illustrates the resulting architecture.

\myparagraph{S3 integration}
\btrlog uses AWS's Rust S3 SDK to flush full log segments to S3.
As \cref{sec:protocol} describes, \btrlog reduces network utilization and costs when flushing to S3:
Log segments flushed to S3 by some node will not be flushed again by another. 
The Rust S3 SDK integrates with Rust's asynchronous ecosystem but exhibits long-running blocking operations (single-digit milliseconds) that deteriorate latency when run on regular log node worker threads.
The implementation thus executes S3 operations on dedicated threads.

\revision[R3.W1 F2]{
  \myparagraph{Concurrent appends}
  The \btrlog protocol permits clients to issue a bounded number of appends concurrently.
  We implement this via an LSN window $W$ respected by both the client library and the log nodes:
  a client does not submit LSN $X$ until a write quorum has acknowledged LSN $X - W$.
  Conversely, log nodes delay flushing segments overlapping with the LSN window to S3, ensuring that the corresponding records remain accessible in memory.
  Beyond allowing pipelining, the LSN window also serves as a retention mechanism for tail-following readers such as database secondaries, as discussed in \cref{subsec:happy_path}.
  A larger window accommodates slower readers but entails higher memory usage per active log stream.
}

\section{Evaluation}%
\label{sec:eval}\label{sec:evaluation}

\myparagraph{Outline}
After detailing the experimental setup, this section evaluates \btrlog across a number of dimensions, including append latency (Sec.~\ref{sec:eval-latency-throughput}), single-AZ and multi-AZ deployment (Sec.~\ref{sec:eval-cross-az}), impact of node failure and quorum (Sec.~\ref{sec:eval-latency-failure}), cost and availability (Sec.~\ref{sec:eval-cost}), and end-to-end OLTP performance (Sec.~\ref{sec:eval-leanstore}).

\myparagraph{Comparison systems}
As discussed, Apache BookKeeper is the only alternative log system that eagerly assigns LSNs and executes appends with a single network round trip.
Since other log systems have higher latency by design, our comparison focuses on Apache BookKeeper.
We also compare \btrlog with Amazon EBS~\cite{Barr_2008} in its low-latency, high-durability variant \textit{io2}, which is a common choice for attaining durable writes on ephemeral cloud instances~\cite{SapEBS}.

\myparagraph{Experimental setup}
We evaluate all systems on a cluster of three \textit{c6id.metal} log nodes with local SSDs in the AWS \textit{eu-central-1} region, plus a \textit{c6in.metal} client node, which has enough network bandwidth to serve all log nodes simultaneously.
The cluster uses ``partitioned'' placement groups, as recommended by AWS~\cite{aws_placement_groups} for high-availability services.
All measurements for an experiment are executed on the same cluster to ensure comparability.
All nodes run Linux 6.14, and \btrlog uses instance-local SSDs as block devices.

\myparagraph{BookKeeper setup}
For BookKeeper, we measured both the default configuration and an optimized configuration for low-latency writes on SSDs.
The default BookKeeper configuration buffers appends for up to \ms{2} before group-flushing them to SSD\@.
This buffering time is a fraction of typical HDD latency, but 60 times the write latency of a modern SSD\@.
Our SSD-optimized BookKeeper configuration removes buffering entirely, aligns writes to \kib{4}, and disables the page cache.
We combine SSDs in a RAID0 using the XFS file system, which has little overhead~\cite{dana,umami}.

\myparagraph{Amazon EBS}
EBS experiments directly attach an \textit{io2} volume to the client node to achieve IOPS comparable with the \btrlog cluster.
This volume is used as a block device, without a file system, similar to how \btrlog log nodes write to their local SSD\@.

\subsection{Append Latency and Throughput}%
\label{sec:eval-latency-throughput}

\begin{figure}
  \centering
  \includegraphics[width=0.95\linewidth, trim={0.1cm 0.2cm 0.1cm 0cm}, clip]{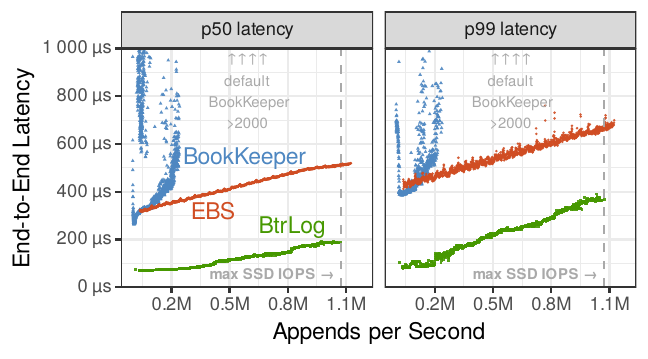}
  \vspace{-1em}
  \caption{Latency with increasing load in EBS (\includegraphics{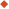}), SSD-optimized BookKeeper (\includegraphics{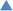}), and \btrlog (\includegraphics{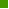}).
  }\label{fig:cmp-latency-throughput}
\end{figure}

\myparagraph{Open-loop latency under load}
Response latency depends on both system configuration and load.
For example, systems may batch disk writes from multiple requests to improve peak throughput at the cost of higher latency.
Therefore, we vary the system load (appends per second), and record the end-to-end latency of each append.
For comparability, we configure all systems to achieve the best possible latency by deactivating write batching.
Our open-loop benchmark schedules requests with exponentially distributed inter-arrival times per log stream, i.e., as a Poisson point process.
The client node adds more independent log streams over time, \revision[F2]{
  increasing system load and stress-testing many concurrent log streams.
}

\myparagraph{Latency with increasing load}
\Cref{fig:cmp-latency-throughput} shows the median and 99th-percentile response latency for BookKeeper, EBS io2, and \btrlog with 128-byte appends, \revision[R3.D4]{which we choose based on observed YCSB, TPC-C, and \textit{pgbench} LSN write sizes.}
Each point corresponds to a \ms{500} interval over which throughput and latency percentiles are measured.
BookKeeper does not scale beyond 240{,}000 appends per second in either configuration (\pointref{bookkeeper-lowlat}, \pointref{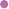}), and its throughput and latency fluctuate strongly under load, producing the chaotic pattern in \Cref{fig:cmp-latency-throughput}.
The default BookKeeper configuration is not visible in the figure because its latency never falls below \ms{2}.
EBS io2 (\pointref{ebs}) shows steadily increasing write latency as load rises and consistently exhibits 4--5\texttimes{} higher median latency than the \btrlog prototype.
At approximately 1\,M appends/s, which is the IOPS limit of the instance-local SSD, both EBS and \btrlog latency deteriorate.
\revision[F3]{
    Additional functionality such as authentication and encryption can be comfortably supported by \btrlog's remaining CPU cycles; stronger hardware-based protection using enclaves~\cite{DBLP:journals/dbsk/LutschEIB25,DBLP:conf/damon/LutschFEIB25,DBLP:conf/damon/El-Hindi0HLZB22} is also an interesting avenue for future work.
}

\myparagraph{Best-case latency}
Let us now discuss interesting latency and system load results for each system using the data points shown in \cref{fig:cmp-latency-throughput}.
Across a large set of BookKeeper configurations, the best median latency we observed was \us{262} at 6{,}800 appends per second; latency deteriorates quickly as load increases.
The best median latency on EBS was \us{318}, which increased steadily with higher load.
\btrlog's best median latency is \us{70} (\us{79} p99 latency) at a load of 35,500 appends per second.
At half of the maximum system load, approximately 500\,k appends per second, \btrlog achieves a median latency of \us{111}.
\revision[F1, F2]{%
  This headroom would also allow adding a \btrlog gateway that exposes a simple authenticated HTTP or gRPC API.
  Even with an additional \us{40} of latency, such a design would remain substantially faster than EBS and BookKeeper.
}

\myparagraph{Full system load}
At the maximum load of 1\,M appends/s (128-byte), EBS reaches \us{503} median latency and \us{651} p99 latency.
Once the provisioned-IOPS volume is saturated, EBS latency increases beyond the range shown in \Cref{fig:cmp-latency-throughput}.
\btrlog reaches \us{188} median latency at full load, limited by the 1.1\,M IOPS of the underlying SSDs.
To enable overload detection and to bound the latency of admitted requests, \btrlog limits its internal queue lengths and drops excess requests once the system is saturated~\cite{hotos/IsaacsAMKRSS25}.
As a result, it does not exhibit the typical ``hockey stick'' latency curve near full load.
\revision[F2]{
  At peak throughput, \btrlog serves 2{,}186 concurrently active logs in this experiment, and log nodes use \gib{71} of main memory.
  While the maximum append rate is bounded by SSD IOPS, the number of logs stored is bounded by main memory because each log consumes \mib{32} on average.
  A production implementation should therefore flush segments of logs with very low append frequency to SSD or S3.
}

\myparagraph{Impact of SSD latency}
Applications with weaker durability requirements may run \btrlog with SSD writes disabled on log nodes:
Each log node appends to log segments in memory and flushes them to S3 when full, but does not write its own WAL to SSD\@.
This setting improves latency, since SSD write latency is similar to the network round-trip time in a single availability zone (\us{35} vs \us{40}).
We find that disabling SSDs reduces median latency by approximately 50\% and improves tail latency across most throughput settings.

\revision[\\R2.D1,\\R3.W3]{
  \myparagraph{Comparison with EBS}
  The large latency gap between EBS and \btrlog has a plausible architectural explanation:
  EBS provides durability using chain replication~\cite{nsdi/BrookerCP20}.
  As discussed in \Cref{subsec:logs_systems}, chain replication requires more network hops than \btrlog's client-driven quorum replication on the append path.
  More broadly, remote block storage exposes a general-purpose block device abstraction rather than an append-only interface and therefore cannot optimize as aggressively for small WAL appends.
  To assess whether this behavior is specific to AWS, we also measured public block storage services on GCP and Azure and compared them to the corresponding network-plus-local-SSD baseline (evaluated with \textit{sockperf} and \textit{fio} using \textit{n4-standard-2} + \textit{hyperdisk-balanced} in GCP \textit{europe-west3}, and \textit{Standard\_D4s\_v3} + \textit{PremiumV2\_LRS} in Azure \textit{germanywestcentral}).
  We find that all hyperscaler block storage services exhibit high write latency compared to the optimum:
  \begin{center}
    \begin{tabular}{l|rrr} \toprule
                                   & AWS           & GCP           & Azure         \\ \midrule
      Network + SSD                & \us{76}       & \us{71}       & \us{301}      \\
      Remote Block Storage Service & \us{311}      & \us{453}      & \us{776}      \\\midrule
      Factor                       & 4.1\texttimes & 6.4\texttimes & 2.6\texttimes \\
\bottomrule
    \end{tabular}
  \end{center}
  These measurements suggest that block storage services in general are not optimized for low latency WAL appends.

  \myparagraph{Comparison with BookKeeper}
  Unlike EBS, BookKeeper and \btrlog share the same replication mechanism: both use client-driven quorum replication and can acknowledge appends in a single round trip.
  To assess whether their remaining latency gap is primarily due to language-level effects, we conducted microbenchmarks showing that Java's networking and SSD I/O primitives can achieve latency comparable to their Rust and C counterparts.
  We therefore attribute the gap mainly to implementation choices in BookKeeper, which was designed around millisecond-scale HDD latency rather than modern low-latency networks and NVMe SSDs.
}

\subsection{Impact of Node Failure and Quorum}\label{sec:eval-latency-failure}

\begin{figure}
  \centering
  \includegraphics[width=\linewidth, trim={0cm 0.2cm 0cm 0cm}, clip]{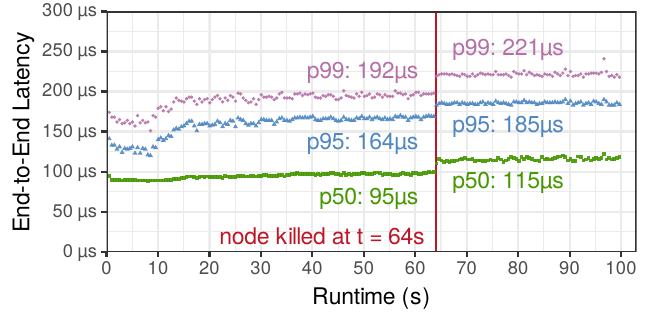}
  \vspace{-1.6em}
  \caption{End-to-End latency before and after killing one log node at 400,000 log appends per second.}\label{fig:node-failure-latency-impact}
\end{figure}

\myparagraph{Impact of node failure}
Our experimental setup with three log nodes can tolerate one node failure, since a quorum of two nodes suffices for appends.
The following experiment tests the append availability and latency of the \btrlog prototype under node failure by manually stopping a random log node while the benchmark is running.
The load is kept constant at approximately 400,000 appends per second.
As \cref{fig:node-failure-latency-impact} shows, the client keeps successfully appending log entries on the \btrlog cluster after a node is killed at $t = 64s$.
The median latency deteriorates from \us{95} to \us{115} (21\%), and the p99 latency from \us{192} to \us{221} (15\%).

\myparagraph{Discussion: quorum latency}
\btrlog's quorum-based algorithm does not need to wait for all responses to arrive, thereby hedging against network latency jitter.
Killing one node effectively stops request hedging and causes the latency increase visible in \cref{fig:node-failure-latency-impact}.

\begin{figure}
  \centering
  \includegraphics[width=0.9\linewidth, trim={0cm 0.2cm 0cm 0cm}, clip]{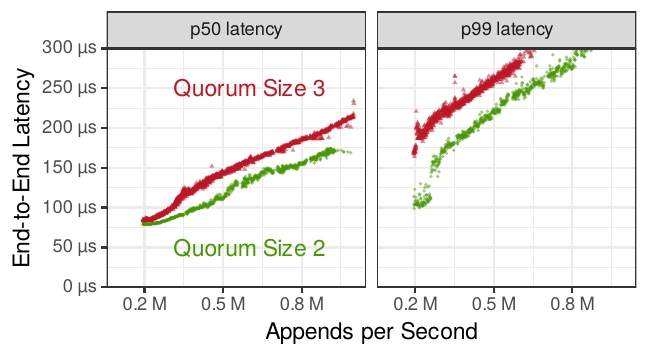}
  \vspace{-.5em}
  \caption{Impact of required quorum (2 or 3 nodes) on median and tail latency.}\label{fig:quorum-latency-impact}
\end{figure}

\myparagraph{Impact of quorum size}
\btrlog can be configured with different quorum sizes.
The default number of responses required for a successful append operation is a majority, e.g., two out of three nodes.
In the following experiment, we investigate the effect of disabling quorum for append requests, so that they require responses from all nodes.
As \cref{fig:quorum-latency-impact} shows, the quorum-based protocol significantly improves end-to-end latency:
At half system load (500,000 appends/s), the quorum improves median latency by approximately \us{30} and p99 latency by approximately \us{50}.

\subsection{Cross-AZ Latency}\label{sec:eval-cross-az}

\myparagraph{Network latency dominates}
Previous experiments used a log cluster inside a single availability zone (AZ).
This yields better latency and availability comparable to EBS, but may not provide enough availability for some use cases.
To validate that \btrlog, like in a single AZ, achieves latency close to the hardware limit, we now spread the three-node cluster across multiple AZs of the same region (eu-central-1).
In this configuration, the client sees idle-load median network latency of \us{270}, \us{440}, and \us{540} to the three log nodes as measured using sockperf~\cite{sockperf}.
Note that these numbers may vary by region.
As the following figure shows, the overall append request latency in \btrlog is indeed dominated by cross-AZ network latency (400,000 appends per second):

\vspace{.5em}
\begin{center}
  \centering
  \includegraphics[width=\linewidth]{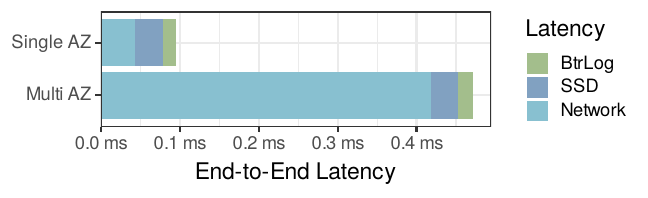}
\end{center}

\noindent 
As the previous single-AZ experiments showed, \btrlog achieves 4\texttimes{} better latency than EBS io2.
In a multi-AZ configuration with three nodes, \btrlog achieves comparable latency (12\% worse) to a single-AZ provisioned EBS volume at the same throughput while providing multi-AZ availability and a WAL-optimized interface.

\subsection{End-to-End Database Performance}\label{sec:eval-leanstore}

\myparagraph{Integration in LeanStore}
Experiments so far have used a custom open-loop benchmarking client designed to measure response latency as accurately as possible under different load settings.
We now turn towards evaluating the performance of these systems in a transactional database, LeanStore~\cite{icde/LeisHK018}.
LeanStore's autonomous commit protocol~\cite{pacmmod/NguyenAZL25} uses multiple log streams (one per worker thread) that are flushed independently.
This design was conceived for local SSDs, which require many parallel writes to saturate bandwidth, but it also fits the BookKeeper and \btrlog APIs, since their multi-tenant architecture supports multiple independent log streams.
We modify LeanStore by swapping its WAL implementation with different backends:
Custom-built backends for BookKeeper and \btrlog, and the default block-device backend for EBS.

\begin{figure}
  \centering
  \includegraphics[width=\linewidth, trim={0cm 0.2cm 0cm 0cm}, clip]{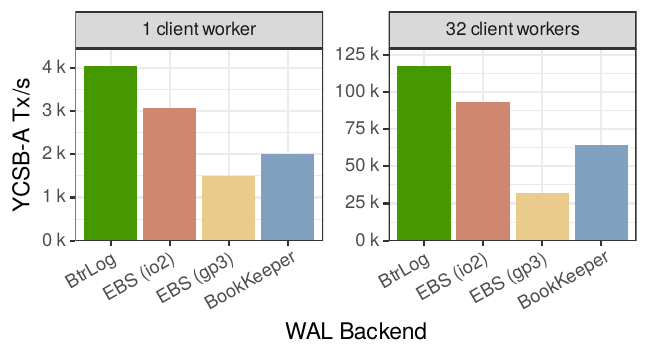}
  \caption{YCSB-A transaction throughput with different WAL backends in LeanStore.}\label{fig:leanstore-throughput}
\end{figure}

\myparagraph{Transaction Throughput}
\cref{fig:leanstore-throughput} shows the measured YCSB-A (50\% reads, 50\% writes) throughput using different WAL backends in LeanStore.
LeanStore using the \btrlog WAL backend achieves 2\texttimes{} higher throughput than LeanStore using the BookKeeper backend, 3\texttimes{} higher throughput than EBS gp3, and 1.25\texttimes{} higher throughput than EBS io2.
These results show that systems previously using EBS io2 for their WAL can replace EBS with a non-provisioned, multi-tenant service while simultaneously increasing transaction throughput using a simple \textit{append} interface.

\revision[\\R1.W1 R1.D3 R2.D2]{
\subsection{WAL Backend Tradeoffs}\label{sec:eval-cost}
Choosing a cloud WAL backend requires balancing append latency with cost and availability.
While the previous results focused on latency, \Cref{fig:system_cost_latency} also compares cost and fault-tolerance for \btrlog and existing alternatives across all three dimensions.

\myparagraph{Append latency and availability}
We measure the \kib{1} append latency using each backend's native append interface in microbenchmarks.
In contrast to \cref{sec:eval-latency-throughput}, we optimize for throughput rather than latency to compare per-append costs:
All systems batch four \kib{1} appends into one \kib{4} I/O.
Corfu, Scalog, and BookKeeper benchmarks use the implementations provided by each GitHub repository~\cite{corfu_repo,scalog_repo,bookkeeper_repo};
S3 Standard and S3 Express benchmarks use the AWS C++ S3 SDK;
EBS gp3 and io2 benchmarks use \textit{fio}~\cite{fio_repo}.
Corfu and Scalog run in a \gbit{400} lab environment; other measurements use AWS \textit{c6id} instances in \textit{eu-central-1}.

\myparagraph{Availability}
Instance-local NVMe SSD writes on EC2 take approximately \us{35}, but provide neither durability nor availability under instance failure.
\Cref{fig:system_cost_latency} thus classifies them as low availability (red).
Deployments replicated across nodes within a single AZ provide higher availability (orange), while deployments spanning multiple AZs, including S3 and the corresponding variants of \btrlog and BookKeeper, provide the highest availability (green).

\myparagraph{Cost}
To compare services with different pricing models, we normalize all costs to \textit{per-append} cost, shown on the x-axis of \Cref{fig:system_cost_latency}.
Cloud-native services such as S3 and S3 Express already use per-operation pricing.
For provisioned services (EBS) and hosted systems such as Apache BookKeeper and \btrlog, we derive per-append cost assuming full resource utilization.

\myparagraph{Detailed cost calculation}
For EBS, we assume peak provisioned performance (256{,}000 IOPS for io2 and 80{,}000 IOPS for gp3) and consider only IOPS cost, excluding storage-capacity cost.
Provisioning 256{,}000 IOPS on EBS io2 costs \$9{,}651.20 per month, or, assuming a 30-day month, \(\$\,1.45 \times 10^{-8}\) per I/O operation.
With full IOPS utilization, this corresponds to \$0.0145 per 1\,M I/O operations and thus \$0.0036 per 1\,M appends.
For hosted systems, including \btrlog, Apache BookKeeper, Corfu, and Scalog, we use the same IOPS-based calculation and assume \textit{c6id.metal} instances, each providing 1{,}073{,}336 IOPS across four local SSDs.
Under this model, single-AZ \btrlog yields a cost of \$0.00125 per 1 million appends, while its additional S3 PUT cost is negligible because appends are batched before upload (for 16\,MB batches and 1\,KB requests, the S3 PUT cost is approximately \$\,$3 \times 10^{-10}$ per append).
Apache BookKeeper incurs higher cost because it dedicates SSD devices to read operations, reducing the IOPS available for writes.
For multi-AZ deployments of \btrlog and BookKeeper, we additionally account for the larger deployment size (six rather than three nodes) and cross-AZ network transfer cost, again assuming 1\,KB requests.

\begin{figure}
  \centering
  \vspace{-.5em}
  \includegraphics[width=\linewidth, trim={0cm 0.2cm 0cm 0cm}, clip]{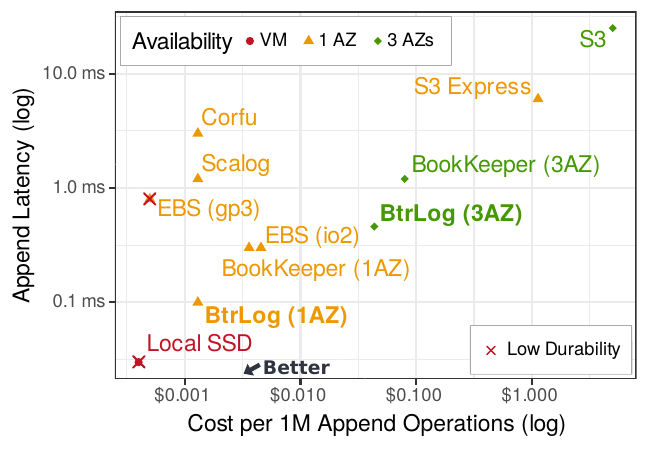}
  \caption{Comparison of append cost, append latency, and availability of potential WAL backends for the cloud. \btrlog shrinks the gap between local SSDs and alternative solutions.}%
  \label{fig:system_cost_latency}
\end{figure}

\myparagraph{Discussion of results}
As \Cref{fig:system_cost_latency} shows (on a logarithmic scale), \btrlog yields the best cost/latency tradeoff in both the single-AZ and multi-AZ availability classes.
Note that these calculations assume the best case for all systems, even those that were not able to fully utilize the SSD IOPS budget in our experiments.
} 

\section{Related Work}

In addition to prior discussions, we review related work on cloud-native databases, shared-log designs, and cloud storage backends.

\myparagraph{Disaggregated and cloud databases}
Decoupling and optimizing logging functionality is a common pattern in modern cloud database systems.
Amazon Aurora~\cite{sigmod/VerbitskiGSBGMK17} uses a log-centric design with a highly coupled log and storage component to reduce network traffic and improve fault tolerance.
Microsoft Socrates~\cite{sigmod/AntonopoulosBDS19} implements a dedicated XLOG service with quorum-based commit, and Huawei TaurusDB~\cite{sigmod/DepoutovitchCCL20} uses distinct log nodes (PLog) to reduce commit latency.
OceanBase PALF~\cite{pvldb/HanLCJZTYXTTWXY24} and FoundationDB~\cite{sigmod/ZhouXSNMTABSLRD21} similarly utilize a dedicated log component in their design.
Neon separates PostgreSQL compute from a replicated WAL service (``safekeepers'') that provides durable WAL ingestion and supports failover~\cite{NeonDBArchitecture}.
Many of these systems, e.g., Aurora, Socrates, OceanBase, and Neon, also integrate object storage for low-cost storage.
Although these systems highlight the need for a specialized logging component, their components are tightly coupled with the database engine and are not described sufficiently in the literature.
In contrast, \btrlog aims to provide a reusable write-ahead logging system that can be integrated with database engines, supporting the development of modular and composable cloud database architectures~\cite{pvldb/Li23,pvldb/Yu25}.

\myparagraph{Shared logs and distributed log systems}
Besides industrial systems, log abstractions have also been studied in academic research.
Hyder proposed using a single totally ordered log on shared flash as the backend for OLTP~\cite{cidr/BernsteinRD11}, motivating systems such as Corfu~\cite{nsdi/BalakrishnanMPWWD12}.
However, shared-log systems including Corfu, Delos, and Scalog emphasize multi-writer semantics and throughput rather than low latency~\cite{nsdi/BalakrishnanMPWWD12,osdi/BalakrishnanFSD20,nsdi/0001CZLAR20}.
Milliscale~\cite{milliscale} recently proposed using S3 Express for OLTP to achieve scalable multi-millisecond latency, while \btrlog targets scalable multi-microsecond latency.
LazyLog~\cite{sosp/LuoBHAG24} and SpecLog~\cite{osdi/BhatHLHGA25} explore lazy or speculative binding of records to log positions to reduce append latency, and FuzzyLog~\cite{osdi/LockermanFKSAA018} relaxes ordering to partial orders to increase concurrency.
These works, however, introduce new append semantics in which LSNs are unknown at commit time, making them incompatible with common recovery protocols such as ARIES. 

\myparagraph{Log-centric streaming systems}
There has also been work on distributed log systems for streaming.
LogDevice is a deprecated \emph{distributed log system} developed by Facebook that was designed to support generic record-oriented and append-only use cases such as write-ahead logging for durability and stream processing~\cite{facebook/LogDevice17}.
Kafka~\cite{kreps2011kafka} popularized the replicated commit-log model for high-throughput streaming and introduced tiered storage~\cite{kafaka-tiered-storage} to archive old log segments to object storage.
Streaming-oriented systems such as Apache Pulsar~\cite{middleware/Pulsar2020} and Pravega~\cite{middleware/TinedoJKJ23}, built on top of Apache BookKeeper~\cite{sigops/JunqueiraKR13} as a backend for durability, also provide data offloading to object storage.
\btrlog complements these works by optimizing for databases instead of streaming use cases: it targets the single-writer, low-latency append pattern of database WAL and leverages ARIES-style semantics to minimize commit overhead, while still supporting asynchronous archival to object storage.
Still, the prevalence of the log abstraction in other systems motivates us to explore using \btrlog for other latency-sensitive use cases.

\myparagraph{Low-latency durable commits}
Related work has studied how fast networks and memory can reduce commit latency.
FaRM~\cite{nsdi/DragojevicNHC14} and RAMCloud~\cite{tocs/OusterhoutGGKLM15} replicate logs to remote DRAM to make commits durable with microsecond-scale overhead.
QueryFresh~\cite{pvldb/WangJP17} uses an RDMA-accessible NVRAM log as primary storage to enable fresh reads on standbys.
In contrast to these works, \btrlog aims to achieve low-latency durability on commodity cloud infrastructure.

\myparagraph{Using cloud storage for databases}
Cloud storage characteristics shape database design beyond WAL.
Durner et al. study how to use object storage for analytics despite higher latency and different access patterns~\cite{pvldb/DurnerL023}.
Other benchmarking efforts characterize cloud storage services and database I/O behavior in cloud-native systems~\cite{pvldb/ZhangJTMCJNWLZ23}.
Early work evaluated alternative database architectures on cloud primitives such as EC2/EBS/S3 and quantified their performance implications~\cite{sigmod/KossmannKL10}.
More recent studies characterize cloud database system architecture tradeoffs~\cite{cidr/0001BLB23,cloud-oltp} and the implications of object vs.\ block storage latency/variance~\cite{pvldb/Tan19ChoosingCloudDBMS}.
Instead of adapting database systems to different storage backends, \btrlog provides a specialized WAL storage backend that requires minimal database adaptation and transparently tiers storage for cold data.

\section{Summary \& Future Work}\label{sec:summary}

\myparagraph{Summary}
We present \btrlog, a reusable single-writer logging service purpose-built for cloud-native database systems.
It combines low-latency durable appends with low-cost, highly durable archival on cloud object storage.
Our evaluation shows that \btrlog improves latency by up to 4\texttimes{} relative to the commonly used EBS io2, translating to higher end-to-end transaction throughput in OLTP systems.
Unlike EBS, \btrlog does not require provisioned IOPS or capacity when run as a multi-tenant service.

\myparagraph{Beyond database systems}
Logging is a primitive with a multitude of applications, and although \btrlog is designed as a WAL backend for database systems, we believe its low latency and cost can also be advantageous in other use cases.
Like BookKeeper, \btrlog can be used as a storage backend for streaming systems such as Apache Pulsar.
Using an additional ordering layer, it can also support multi-writer semantics, similar to Corfu and Scalog.

\myparagraph{Towards extensibility}
\btrlog provides a reusable interface and is not coupled to a specific database engine.
One can conceive of further \btrlog extension points in an open ecosystem.
For example, custom log segment filtering and transformation functions on log nodes enable writing to S3 in application-specific formats~\cite{ginter2026active}, or verification metadata~\cite{DBLP:conf/vdbs/El-Hindi0B23} for long-term storage.
This can be achieved securely using WebAssembly functions, which have recently been used for future-proof storage formats~\cite{DBLP:journals/pvldb/GienieczkoKNLG25,DBLP:journals/pacmmod/ZengMPMPPZ25}.
Custom archival backends, such as a page materialization service, enable optimizations for cloud-native OLTP systems, such as Socrates~\cite{sigmod/AntonopoulosBDS19}.

\begin{acks}
    We thank Carsten Binnig for fruitful discussions during the early stages of the project;
    Philipp Unterbrunner, Pat Helland, Anub Ghatage, and Michael Haubenschild for their valuable feedback and industry insights;
    and the anonymous reviewers for their feedback.

  \noindent\includegraphics[width=2em]{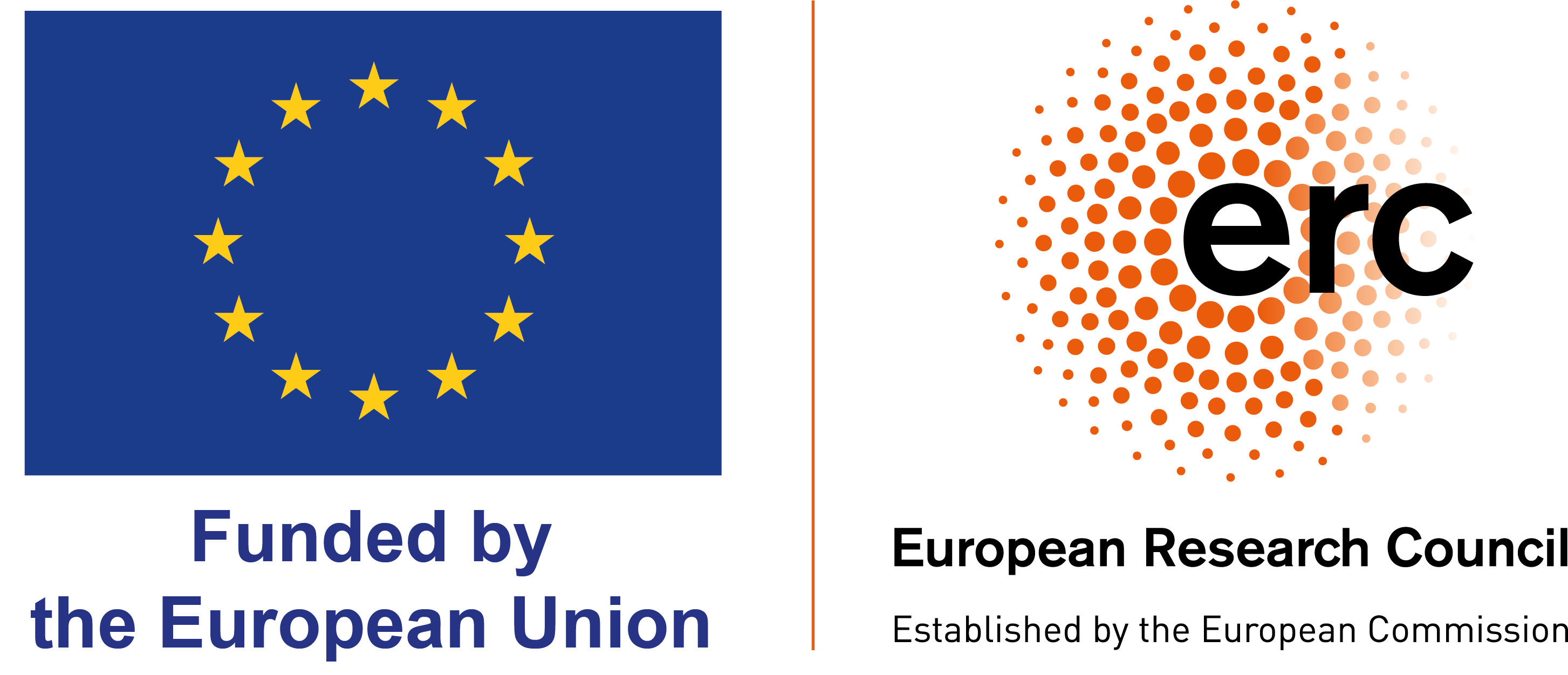} Funded/Co-funded by the European Union (ERC, CODAC, 101041375). Views and opinions expressed are however those of the author(s) only and do not necessarily reflect those of the European Union or the European Research Council. Neither the European Union nor the granting authority can be held responsible for them.
\end{acks}

\balance
\bibliographystyle{ACM-Reference-Format}
\bibliography{references}

\end{document}